\documentstyle[pre,multicol,aps]{revtex}

\input epsf

\begin{document}

\draft

\title{{\rm Phys.\ Rev.\ E Rapid Comm., submitted}
\hfill {MA/UC3M/{\rm 04}/{\rm 1995}}\\[5mm]
Soliton pinning by long-range order in aperiodic systems}

\author{Francisco Dom\'{\i}nguez-Adame\thanks{e-address:
\tt fimat02@emducms1.sis.ucm.es}}
 \address{Departamento de F\'{\i}sica de Materiales, Facultad
 de F\'{\i}sicas, Universidad Complutense, E-28040 Madrid,
Spain}

\author{Angel S\'{a}nchez\thanks{e-address: \tt anxo@dulcinea.uc3m.es}}
 \address{Escuela Polit\'{e}cnica Superior, Universidad Carlos
III de Madrid, C./ Butarque 15, E-28911 Legan\'{e}s, Madrid, Spain}

\author{Yuri S.\ Kivshar\thanks{e-address: \tt ysk124@rsphy1.anu.edu.au}}
 \address{Optical Sciences Centre, Australian National
University, ACT 0200 Canberra, Australia}

\maketitle

\begin{abstract}

 We investigate propagation of a kink soliton along
 inhomogeneous chains with two different constituents, arranged
 either periodically, aperiodically, or randomly. For
 the discrete sine-Gordon equation and the Fibonacci and
 Thue-Morse chains taken as examples, we have found that the
 phenomenology of aperiodic systems is very peculiar:  On the
 one hand, they exhibit soliton pinning as in the random chain,
 although the depinning forces are clearly smaller. In
 addition, solitons are seen to propagate differently in the
 aperiodic chains than on periodic chains with large unit
 cells, given by approximations to the full aperiodic sequence.
 We show that most of these phenomena can be understood by
 means of simple collective coordinate arguments, with the
 exception of long range order effects. In the conclusion we
 comment on the interesting implications that our work could
 bring about in the field of solitons in molecular
 (e.g., DNA) chains.

\end{abstract}

% Quasicrystals

\begin{multicols}{2}

\narrowtext

 The subtle interplay between nonlinearity and disorder is
 being laboriously unveiled throughout the past few years
 \cite{reviews}. A rich diversity of phenomena stems from such
 interaction, their manifestations being found in a number of
 systems ranging from condensed matter physics to biophysics
 \cite{procs}. A number of models have been set forth which
 capture the essential ingredients of those systems while
 enjoying a canonical, non-specific view of the problem. Among
 the most successful of these models, the sine-Gordon
 (SG) equation is particularly remarkable both for its range of
 applicability and the possibilities it opens for study either
 in continuous or discrete version. Some of the physical
 situations well modeled by this equation are, for instance,
 Josephson junctions \cite{BP}, Josephson junction arrays
 (JJA) \cite{JJA,Matteo}, or DNA promoter dynamics
 \cite{Salerno,YS}. Importantly, many realistic systems like
 DNA chains are neither periodic nor random, being inherently
 close to quasi-periodic or aperiodic systems, so that the
 effects of long-range order may change the dynamics of
 nonlinear excitations.

 In this Rapid Communication we concern ourselves with the
 problem of the behavior of kink solitons on lattices
 consisting of two different components, thereby focusing on
 issues inherently discrete similar to those of DNA or JJA
 dynamics. Our main aim here is to learn about {\em the
 phenomenology of soliton propagation as a function of the
 order of the underlying lattice}. We consider three main
 possibilities for our binary chain: {\em periodic},
 {\em aperiodic}, and {\em random}, which represent,
 respectively, full order, long-range order, and pure disorder.
 We show in the following that, while the periodic lattice
 exhibits basically the same features as the homogeneous case,
 the two non-periodic systems present characteristics of their
 own. We further discuss how most of our results can be
 understood within the framework of the collective coordinate
 technique \cite{MS} (see also the review \cite{YB} and
 references therein).  Notwithstanding that analytical insight,
 we have also found effects that cannot be interpreted in terms
 of such a particle-like behavior, and we have been able to
 associate those to the long-range order characteristics of aperiodic
 chains.

 The model we use as our working example is a damped, dc
driven, discrete SG equation given by \begin{equation}
\label{dsg}
 \ddot{u}-\,{1\over a^2}(u_{n+1}-2u_n+u_{n-1})+V_n\sin u_n+
\alpha\dot{u}=F, \end{equation} where dot means time
derivative, $a$ is the lattice spacing, and $n$ runs over the
 lattice sites $n=1,\ldots, N$. The coefficient in front of the
 on-site potential, $V_n$, is directly related to the physical
 properties of the application one is interested in: Thus, it
 has to do with local critical currents in Josephson devices,
 or with the strength of hydrogen bonds between complementary
 basis in DNA models. In the following, we will allow $V_n$ to
 take only on two values, $V_a$ and $V_b$. Moreover, by an
 appropriate rescaling, it is possible to fix $V_a=1$, and so
 will be done hereafter. The spatial arrangement of the two
 kinds of values will be chosen to be either periodic,
 aperiodic, or random. As our aperiodic models, we pick two
 standard choices, namely the Fibonacci and the Thue-Morse
 chains. They are generated starting from two basic units $A$
 and $B$ using the following inflation rules: $A\to AB$, $B\to
 A$ for the Fibonacci chain and $A\to AB$, $B\to BA$ for the
 Thue-Morse chain. In this way, finite and self-similar
 aperiodic chains are obtained by $n$ successive applications
 of these rules, with $N=F_n$ sites for the Fibonacci lattice
 and $N=2^n$ sites for the Thue-Morse lattice. Here
 $F_n=F_{n-1}+F_{n-2}$ with $F_0=F_1=1$ are the Fibonacci
 numbers. The number of $A$-sites in the lattice is $\sim \tau
 N$ in the Fibonacci case and $N/2$ in the Thue-Morse case,
 where $\tau=\lim_{n\to \infty} (F_{n-1}/F_n)=(\sqrt{5}-1)/2$
 is the inverse golden mean. Both chains have been used in very
 many contexts to model aperiodic ordering which, in spite of
 previous, more naive ideas, it is not something intermediate
 between periodic and random systems (see, e.g., \cite{us} and
 references therein).

\begin{figure}
\setlength{\epsfxsize}{6.0cm}
\centerline{\mbox{\epsffile{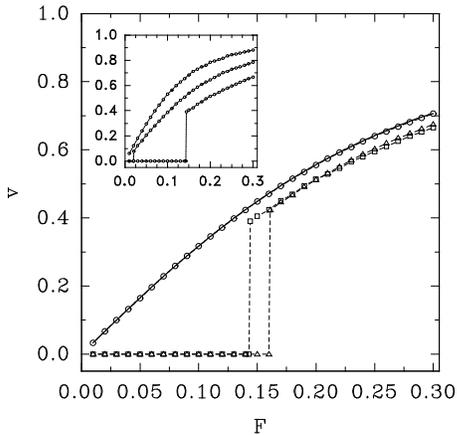}}}
\caption{Steady state velocity {\em versus} applied force for
chains with different orderings and $V_b=10$.
Points correspond to numerical
results, solid lines to the theoretical prediction in
Eq.~(\protect\ref{speed2}) with $V_{\em avg}$
corresponding to $c=0.5$. ($\bigcirc$), periodic
chain with $N=4\,182$; ($\Box$), Fibonacci chain with
$N=F_{18}=4\,181$; ($\bigtriangleup$), Thue-Morse chain with
$N=2^{12}=4\,096$. Dashed lines joining symbols are a guide to
the eye. The threshold for random chains is out of the range of
the plot and its value is about $F=0.5$. Inset: same but for
Fibonacci chains with $V_a=2$, $V_a=5$, and $V_a=10$, from top
to bottom.}
\label{fig1}
\end{figure}

 To characterize the dynamics of kink solitons on these
systems, we have numerically simulated Eq.\ (\ref{dsg}) by
means of a fifth-order adaptive-stepsize Runge-Kutta routine
\cite{Recipes} which has been shown to be an accurate procedure
(see \cite{psg} and references therein). In the homogeneous
case ($V_n=V_a=1$), it has long been known \cite{MS} that if a
soliton initially at rest evolves according to Eq.\ (\ref{dsg})
with $F>0$, it eventually reaches a steady state in which it
propagates along the chain with velocity \begin{equation}
\label{speed}
 v=\left[1+\left(\frac{4\alpha}{\pi F}\right)^2\right]^{-1/2},
\end{equation} with the opposite sign if $F<0$. This result was
found by means of an standard collective coordinate
calculation, and the derivation can be found in \cite{MS}. As
another check of our simulations, we compared their outcome to
this prediction and found an agreement better than $1\%$ for
all studied cases. We will also make use of this expression,
although in a modified form: If one assumes $V_n$ is constant
and given by the average of $V_a$ and $V_b$ weighted by their
concentrations, say $V_n=V_{\em avg}=cV_a+(1-c)V_b$ (here $cN$
is the number of $A$-sites of the chain), and repeats the same
calculation in \cite{MS}, the predicted final velocity is
\begin{equation}
\label{speed2}
v=\left[1+V_{\em avg}
\left(\frac{4\alpha}{\pi F}\right)^2\right]^{-1/2}, \end{equation}
 i.e., the asymptotic velocity is predicted to be smaller
(higher) than that of the homogeneous model when $V_b>V_a$ ($V_b<V_a$).

 The results of our numerical simulations are collected in
Fig.\ \ref{fig1}, where we present the value of $v$ as a
function of the applied force $F$ for the different orderings
considered, always with $V_b=10$.
The plotted value of $v$ was obtained by starting
the simulation for each $F$ with a kink at rest in the middle
of the chain, as given by the exact continuum solution of the
SG equation, and letting it evolve while monitoring
its velocity until it reached a constant value. In all cases it
was verified that the soliton shape remained almost unaltered,
which is the necessary condition for the concept of velocity to
make sense. The simulation parameters are $\alpha=0.1$, which
is a prototypical value and whose only effect is to fix the
force scale, and $a=0.1$, a value which yields a discrete chain
but still close to the continuum to avoid side effects induced
by pinning due to the effective Peierls-Nabarro potential
\cite{PN}.  In Fig.\ \ref{fig1} it is easily appreciated the
different behavior of the different kinds of ordering
considered. The periodic chain with alternating
$V_a$ and $V_b$ is very much accurately described by our
theoretical prediction in Eq.~(\ref{speed2}) with average
velocity $V_{\em avg}$ corresponding to a concentration of $c=0.5$,
whereas it is seen that neither Fibonacci (with $c=\tau$)
nor Thue-Morse chains
obey that equation. The first discrepancy arises as the
existence of a threshold force, $F_{c}$, below which solitons
are pinned and do not move. A comparison of such $F_c$ for the
aperiodic chains with two random chains (not shown in the
figure) with the same concentration of $V_a$
($\tau=0.618\ldots\,$ for the Fibonacci case and $0.5$ for the
Thue-Morse case) leads to the conclusion that the threshold is
about {\em three times higher for the fully disordered chain
than for the corresponding aperiodic one}.  Above threshold,
the approximation (\ref{speed2}) fails also to accurately
predict the value of $v$, overestimating it appreciably for
those non-periodic cases. However, the overall behavior of the
random and the aperiodic chains is different, in the sense that
in the random case, above threshold the $ v-F$ curve is
nonmonotonous and the final velocity depends strongly on the
particular realization of disorder. As for the dependence of
the results on the value of $V_b$, it can be seen in the inset
 of Fig.\ \ref{fig1} that, while keeping $V_b>V_a$, the higher
 $V_b$ the larger $F_c$ for Fibonacci lattices. On the other
 hand, when $V_b=0.1$, i.e., much smaller than $V_a$, we have
 found that, contrary to the naive expectations, there is still
 a threshold force for the Fibonacci case, about $F_c=10^{-3}$,
 which we have verified is not related to Peierls-Nabarro
 pinning because the periodic chain shows no pinning for such
 value of the force. This is a striking result that should be
 compared to the random chain one, where it is found that $F_c
 \simeq 0.04$

\begin{figure}
\setlength{\epsfxsize}{6.0cm}
\centerline{\mbox{\epsffile{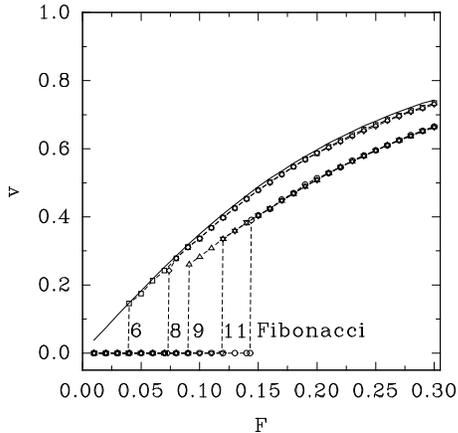}}}
 \caption{Steady state velocity {\em versus} applied force for
chains with different orderings. Points correspond to numerical
results, solid lines to the theoretical prediction in
Eq.~(\protect\ref{speed2}) with $V_{\em avg}$
corresponding to $c=\tau$. ($\Box$), $6\,$th
approximation (13 sites); ($\Diamond$), $8\,$th approximation
(34 sites); ($\bigtriangleup$), $9\,$th approximation (55
sites); ($\bigtriangledown$), $11\,$th approximation (144
sites); ($\bigcirc$), full Fibonacci chain. All lenghts are
close to 4\,180 sites. A soliton at rest spans about 50 sites
for comparison. Dashed lines joining symbols are a guide to the
eye. Labels indicate the order of the approximation to the
Fibonacci chain as well as the full aperiodic
sequence.} \label{fig2} \end{figure}

 To gain further insight into the pinning phenomenon, we have
 compared the results for the full aperiodic chains with those
 of periodic chains with unit cell formed by a shorter
 approximation to the corresponding aperiodic sequence. Thus, for the
 Fibonacci chain, we have studied periodic chains with unit
 cells $V_aV_bV_a$ (the $3\,$th Fibonacci approximation),
 $V_aV_bV_aV_aV_b$ (the $4\,$th approximation), and so forth up
 to the $11\,$th approximation (144 sites in the unit cell).
 The results are shown in Fig.~\ref{fig2}. The most remarkable
 conclusion that can be drawn from this plot is that for
 periodic chains with unit cells smaller than the soliton size
 (around 50 sites when at rest), the behavior above threshold
 is independent of the unit cell, and moreover, it is very well
 described by the average velocity introduced in
 Eq.~(\ref{speed2}). This is very important, since it implies
 the existence of influences coming from the long-range order
 of the full aperiodic chain which do not arise in periodic
 approximations unless the size of the unit cell is much larger
 than the soliton width, i.e., unless the soliton is unable to
 distinguish the unit cell from the whole chain. Another remark
 in order here is that the threshold for the different periodic
 approximations depends on the initial position of the soliton
 in the chain, changing up to a factor two for different
 positions, although keeping below than that of the full
 Fibonacci chain. Results for the Thue-Morse model are
 basically the same, although in this case shorter lengths are
 needed for the simulation to get close to that of the full
 aperiodic chain, and the threshold dependence on the size is
 nonmonotonous. We tentatively associate this to the fact that
 in the Thue-Morse sequence $B$-sites may appear in pairs,
 contrary to the Fibonacci sequence, and therefore it is to be
 expected that their effect will be stronger on the soliton.

\begin{figure}
\setlength{\epsfxsize}{6.0cm}
\centerline{\mbox{\epsffile{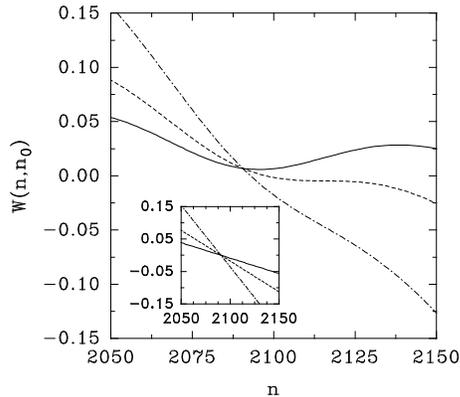}}}
 \caption{Effective potentials for a kink soliton initially at
rest in the center of a Fibonacci chain for forces $F=0.05$
(solid line), $F=0.1$ (dashed line), and $F=0.2$ (dot-dashed
line).  Inset: Same, but for the periodic chain.} \label{fig3}
\end{figure}

 The existence of a threshold force is clearly the main failure
of the collective coordinate theory for both the aperiodic and
the random chains and, therefore, we undertook the task of
finding a better analytical description. To this end, we
followed the same approach of the work by Kivshar and Salerno
\cite{YS}, where they introduced an effective potential to
account for their results on DNA promoter dynamics. The basic
idea is similar to that of the collective coordinate technique,
but they improve it by including the spatial ordering of the
chain. We skip the details here, as the interested reader may
find them in Refs.\ \cite{Salerno,YS}, and quote only the final
result: The effective potential seen by a soliton, initially at
rest at a lattice site $n_0$, is given by
\begin{eqnarray}
 W(n,n_0) & = & \frac{\sum_m(V_{\em avg}+V_n)[{\rm sech}^2(z_m)-{\rm
sech}^2 (z^{(0)}_m)]}{2\sum_m{\rm sech}^2(z_m)}\nonumber\\
\label{ep}
& - & \frac{2F[(\tan^{-1}(e^{z_m})
 -\tan^{-1}(e^{z^{(0)}_m})]}{2\sum_m{\rm sech}^2(z_m)},
\end{eqnarray}
with $z_m\equiv aV_{\em avg}^{1/2}(m-sn)$ and
$z^{(0)}_m\equiv aV_{\em avg}^{1/2}(m-sn_0)$, and the sums run
 over the whole lattice. Finally, to include soliton width
 effects, the potential in Eq.~(\ref{ep}) is averaged in the
 interval of the lattice spanned by the soliton. We have to
 stress that this approach only applies to the early stages of
 the problem, when the kink is at rest or beginning to move at
 a very slow speed. This is so because in deriving
 Eq.~(\ref{ep}) the dissipative term is not included.
 Therefore, this approach should be useful in predicting the
 threshold force although it certainly does not apply
 to the
 dynamics above threshold. As can be seen from Fig.~\ref{fig3},
 the agreement is very good for the Fibonacci chain, and the
 same can be said about the Thue-Morse chain and the random
 case (not shown; in addition, the random case depends strongly
 on the realization considered, which was to be expected in
 view of the simulations). For comparison, notice that the
 potential minimum is always absent in periodic lattices (see
 inset of Fig.~\ref{fig3}), in agreement with our numerical
 simulations where pinning is not observed in those lattices.
 In computing the average value of $W(n,n_0)$ we have used a
 soliton width of 40 sites, which is of course quite arbitrary.
 We have checked that variations of $\pm 10$ sites are not
 crucial for our results, which remain semiquantitatively
 correct.  Indeed, due to this freedom in the election of the
 soliton width, we have not pursued a better agreement, because
 it would be difficult to justify the choice of that value
 aside from the fact that it fitted the numerical simulations.

 In conclusion, we have studied soliton propagation along
 binary lattices with different orderings. We have found that whereas
 solitons can propagate for any force on periodic systems,
 non-periodic ones exhibit a threshold value, i.e., forces
 larger than a value $F_c$ are needed in order to start
 propagation. We have been able to analytically explain that,
 as well as to characterize the differences that, in turn,
 separate the phenomenology of random and aperiodic chains. We
 have also found that long-range order effects arise when
 solitons are propagating along the chain, which show up in a
 decreasing of the steady state velocity with respect to the
 theoretical expectations for the periodic chains. These
 results can be useful in the context of Josephson devices, as
 they can be the basis for the design of new devices with
 specific properties (the value $F_c$ corresponds to
 a critical current for the device to start conducing).  This
 is also important as our results can be checked in
 specifically built JJA's. On the other hand, they can also be
 relevant to DNA promoter dynamics. Indeed, our results show
 that the long-range correlation present in DNA due to the
 information it encodes makes soliton propagation easier than
 if it were purely random, in fact allowing their traveling
 along the chain at lower velocities.  This would not be
 possible if the structure of the molecule would be random in
 view of our results.

 It is clear that these results just opened the door to
 the problem of soliton propagation in aperiodic systems. As we
 have mentioned along the paper, there are a number of unsolved
 questions, like an analytical explanation of the soliton
 velocity in aperiodic chains, or how periodic approximations
 converge to the full aperiodic system. Besides that, the study
 of other aperiodic models would be helpful in order to clarify
 the generic properties exhibited by this kind of models. Work
 along these lines is in progress.

 We thank E.\ Maci\'a for illuminating conversations on
 aperiodic systems and J.\ A.\ Cuesta, F.\ Falo, L.\ M.\
 Flor\'\i a, P.\ J.\ Mart\'\i nez, and P.\ Qu\'emerais for
 helpful discussions, and L.\ Garc\'\i a for the use of his
 computer facilities.

\end{multicols}

\end{document}